\documentclass[journal]{IEEEtran}
\usepackage{textcomp}
\usepackage{cite}
\usepackage{tabularx}

\ifCLASSINFOpdf
\usepackage[pdftex]{graphicx}

\graphicspath{{Figures/}{Resubmission/}}

 \DeclareGraphicsExtensions{.pdf,.jpeg,.png}
\else
\fi
\usepackage{array}

\newcolumntype{M}{>{\centering\arraybackslash}m{\dimexpr.25\linewidth-2\tabcolsep}}

\usepackage{fixltx2e}
\usepackage[usenames,dvipsnames,svgnames,table]{xcolor}
\usepackage{adjustbox}
\begin{document}
	%
	\title{Energy-Efficient Memories using Magneto-Electric Switching of Ferromagnets}
	%
	%
	%
	
	\author{Akhilesh~Jaiswal,
		Indranil~Chakraborty
		and~Kaushik~Roy,~\IEEEmembership{Fellow,~IEEE}
		\thanks{A. Jaiswal, I. Chakraborty and K. Roy are with the School
			of Electrical and Computer Engineering, Purdue University, West Lafayette, IN
			, 47907 USA e-mail: jaiswal@purdue.edu;~ichakra@purdue.edu; ~kaushik@purdue.edu. The work was supported in part by, C-SPIN, a MARCO and DARPA sponsored StarNet center, by the Semiconductor Research Corporation, the National Science Foundation, Intel Corporation and by the DoD Vannevar Bush Fellowship.}
		}

		\vspace{-148mm}
	
	\maketitle

	\begin{abstract}
	
Voltage driven magneto-electric (ME) switching of ferro-magnets has shown potential for future low-energy spintronic memories. In this paper, we first analyze two different ME devices \textit{viz.} ME-MTJ and ME-XNOR device with respect to writability, readability and switching speed. Our analysis is based on a coupled magnetization dynamics and electron transport model. Subsequently, we show that the decoupled read/write path of ME-MTJs can be utilized to construct an energy-efficient dual port memory. Further, we also propose a novel content addressable memory (CAM) exploiting the compact XNOR operation enabled by ME-XNOR device.

	\end{abstract}
	
	\begin{IEEEkeywords}
		Magneto-electric effect, CAM, dual port, memory, XNOR, LLG.
	\end{IEEEkeywords}

	\vspace{-3mm}

	%
	\IEEEpeerreviewmaketitle

	\vspace{-3mm}
	
	\section{Introduction}
	%
	%
	%
	%
	\IEEEPARstart{M}{agneto-resistive} memories based on current driven Spin Transfer Torque (STT) \cite{toshiba}, have attracted immense research interest due to their non-volatility, almost unlimited endurance and area-efficiency \cite{fong_bit}. However, STT based memories suffer from inherent low switching speed and high write-energy consumption \cite{jaiswal_sca}. Recently, voltage induced Magneto-Electric (ME) effect,  has shown potential for fast and energy-efficient switching of ferromagnets \cite{Revival_ME}. 
	
Many device proposals for memory \cite{me_mem_1}, \cite{me_mem_2} and logic applications \cite{me_logic_1,Intel_ME,mesl} of the ME effect can be found in the literature.  In this paper, we explore two different ME devices  - i) ME magnetic tunnel junctions (ME-MTJs) \cite{me_logic_1} and ii) ME-XNOR device \cite{me_logic_1, mesl}. We analyze the ME devices with respect to writability, readability and switching speed using a coupled magnetization dynamics and transport model. Further, we propose two novel energy-efficient memories - i) a dual port memory and ii) a content addressable memory (CAM), using the aforementioned ME devices.

	\vspace{-2mm}
	
	\section{ ME effect}
Various single phase \cite{BFO_story} and composite multi-ferroic materials \cite{strain} have been experimentally demonstrated to exhibit the ME effect. ME effect is due to exchange bias coupling in single phase materials \cite{ramesh_exchange} and is usually due to strain coupling \cite{strain} in case of composite materials. For example,  in single phase BiFeO$_3$ due to the coupling between the ferro-electric polarization, the (anti) ferromagnetism of BiFeO$_3$, and the ferromagnetism of an underlying nano-magnet, the magnetization of the nano-magnet can be switched by application of an electric field \cite{determinis}. Similarly, strain coupled magnetization reversal in PMN-PT has been proposed in \cite{strain} .

Note, since multi-ferroics in general and ME effect in particular, is currently an area of intense research investigation, we do not follow a particular material set or experiment. Rather, in this work, we treat the ME effect by a generic parameter referred to as the magneto-electric co-efficient ($\alpha_{ME}$) \cite{Intel_ME, determinis, mesl} (explained later in the manuscript). Such an abstraction of the ME effect is justified, since the aim of the present paper is not to explore the various physical phenomenons driving the ME effect. Instead we intend to examine the implications of ME based devices with focus on memory applications. 
\vspace{-3mm}

\section{ ME devices under consideration}
\vspace{-5mm}

		\begin{figure}[h!]
			\centering
			\includegraphics[width=3.3in,keepaspectratio]{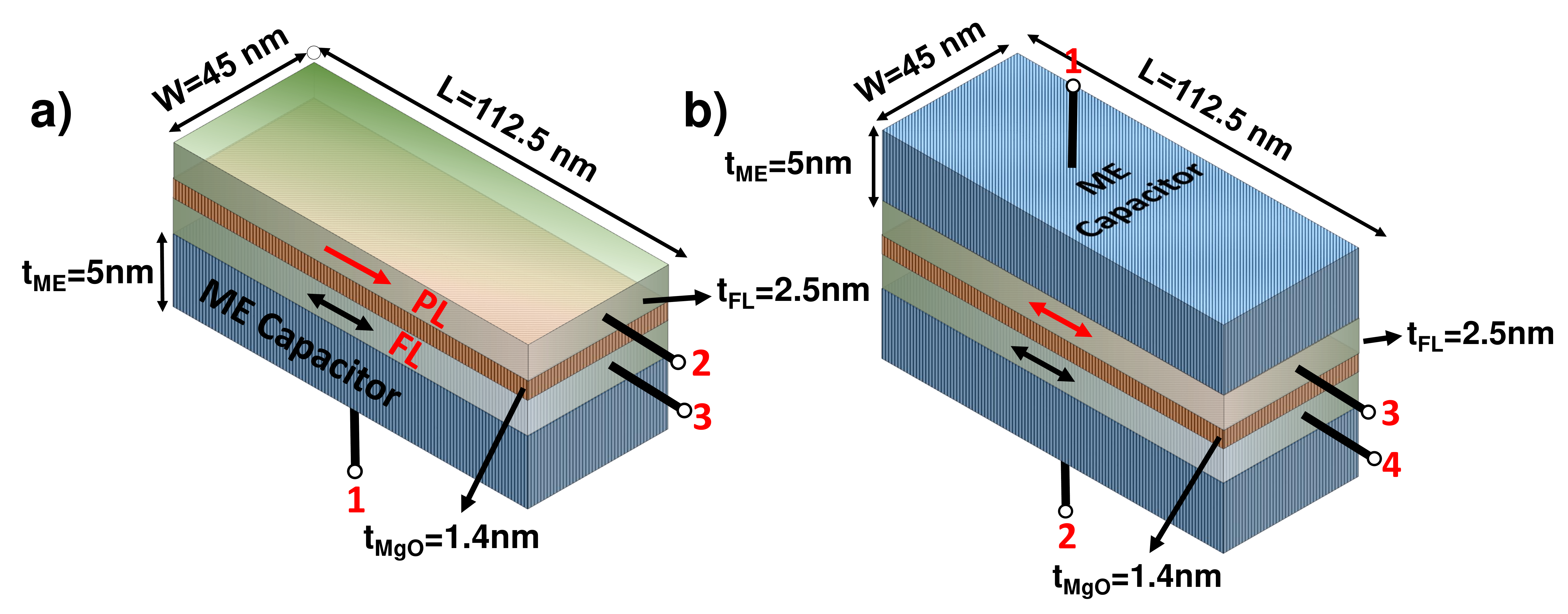}
			\vspace{-4mm}
			\caption{(a) Schematic of the ME-MTJ and (b) ME-XNOR. We have assumed $M_S = 1257.3 KA/m$ \cite{ikeda}, $\alpha  =  0.1$, $K_i = 1mJ/m^2$ \cite{ikeda}, $\alpha_{ME} = 1/c~ms^{-1}$, where c is speed of light, $\epsilon_{ME} = 500$ \cite{Intel_ME}, $T=300K$.}
				\label{ME_devices}
		\end{figure}
			\vspace{-2mm}
We consider two ME based devices $-$ ME-MTJ \cite{me_logic_1} and ME-XNOR \cite{me_logic_1, mesl}, with focus on memory applications. ME-MTJ consists of an MTJ in contact with an ME oxide underlayer as shown in Fig. \ref{ME_devices}(a). The MTJ itself is composed of a \textit{pinned layer} (PL), a \textit{free layer} (FL) and an oxide spacer (usually MgO \cite{why_mgo}). Depending on the orientations of the free and the pinned layer the ME-MTJ can be in either low resistance parallel (P) state or high resistance anti-parallel (AP) state.  The normalized difference in the resistances of the AP and P state is expressed by the tunnel magneto-resistance (TMR) ratio of the MTJ.

In order to switch the ME-MTJ from P (AP) to AP (P) state a positive (negative) voltage exceeding a certain threshold needs to be applied on terminal 1 in Fig. \ref{ME_devices}(a). 
The metal contact to the ME oxide, the ME oxide itself and the free layer of the MTJ can be considered as a capacitor. 
 On the other hand, the value stored in the ME-MTJ can be read by sensing the resistance between terminals 1 and 2. 

In Fig. \ref{ME_devices}(b) we show the ME-XNOR device. The ME-XNOR device consists of two free layers separated by MgO and in contact with respective ME oxides. If the voltage polarity on the terminals 1 and 2 are the same, the MTJ stack would be in P state (measured between terminals 3 and 4), while a different voltage polarity on the two terminals would lead to an AP state. Thus, the proposed device emulates an XNOR functionality. ME-XNOR device in previous works have been used for logic applications \cite{mesl}. In this work, we would later show that ME-XNOR device can be used to construct an energy efficient CAM. In the next section, we describe the simulation model. 
		
			\begin{figure}[t]
			\centering
			\vspace{-5mm}
			\includegraphics[width=3in,keepaspectratio]{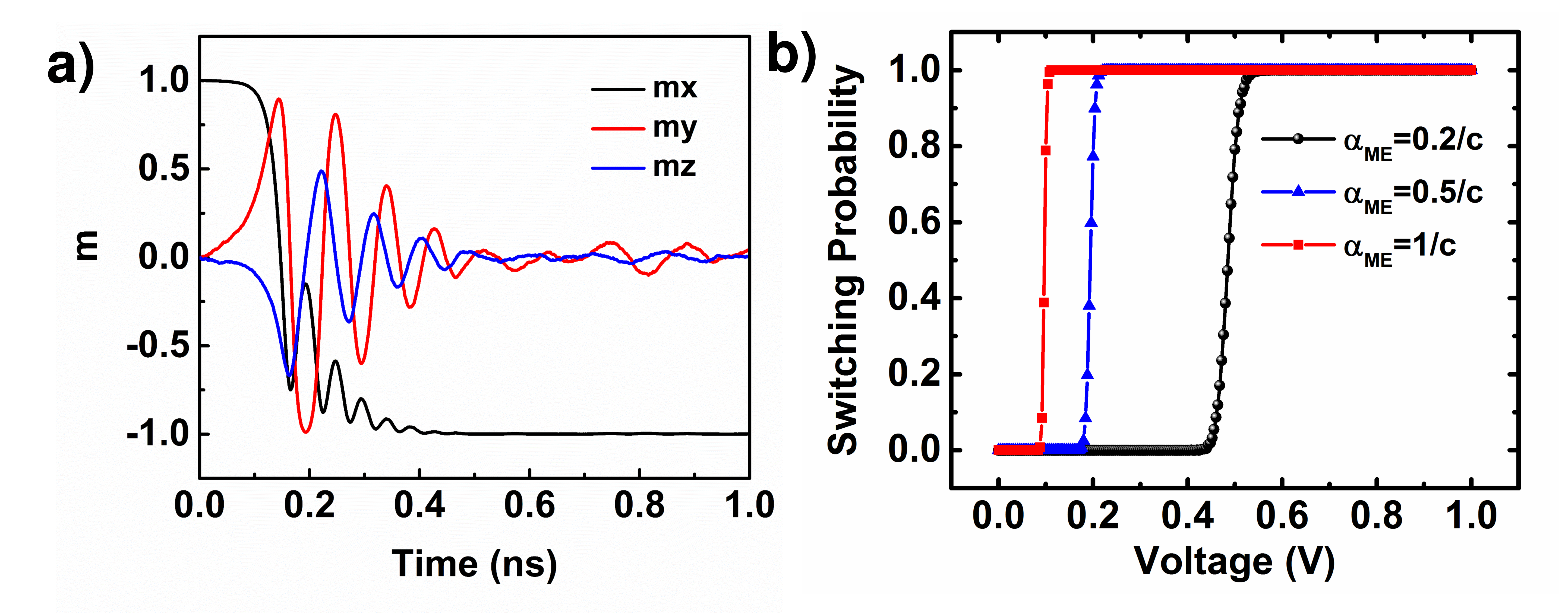}
			\vspace{-4mm}
			\caption{(a)  Magnetization dynamics. (b) Switching probability versus voltage.}
			\label{fig2}
			\vspace{-6mm}
		\end{figure}

	\vspace{-4mm}
	\section{Device Modeling}	
	\vspace{-2mm}
	
	 Under mono-domain approximation, magnetization dynamics can be modeled using the LLG equation, proposed by \textit{Landau, Lifshitz and Gilbert}, as shown below \cite{llg}, \cite{llg_thesis} 
	 	
		\vspace{-3.4mm}
	\begin{equation}
	\frac{\partial \hat{m}}{\partial t}=-|\gamma|  \hat{m}\times H_{EFF}+\alpha \hat{m}\times\frac{\partial  \hat{m}}{\partial t}
	\end{equation}
			\vspace{-1mm}
where $H_{EFF}$ is the effective magnetic field. $H_{EFF}$ is the sum of
	the demagnetization field \cite{demag}, \cite{demag_f}, the interface anisotropy field \cite{jaiswal_sca} and any other external field. $\hat{m}$ is the unit magnetization vector, $\gamma$ is the gyromagnetic ratio and $\alpha$ is the Gilbert damping constant.  The thermal noise is modeled using the Brown's model \cite{Brown} and is accounted for by expressing a contributing field to $H_{EFF}$ as  $\vec{H}_{thermal} = \vec{\zeta} \sqrt {\frac{2\alpha kT}{|\gamma|M_SVdt}}$,
	where $\vec{\zeta}$ is a vector with components that are zero mean Gaussian random variables with standard deviation of 1. $V$ is volume of the free layer, $T$ is the temperature and $k$ is the Boltzmann's constant and $dt$ is time step. The ME effect can be included in $H_{EFF}$ by writing the ME field as \cite{Intel_ME} $H_{ME} = \frac{1}{\mu_0}\alpha_{ME}E = \frac{1}{\mu_0}\alpha_{ME}\frac{V_{ME}}{t_{ME}}$, where the magneto-electric constant is $\alpha_{ME}$ \cite{nikonov_bench}, $E$ is the electric field and $V_{ME}$ is the voltage across the ME capacitor.  
	
	Equation (1) can be solved numerically through the Heun's method \cite{Heun}. In addition, we used the Non Equilibrium Green's Function (NEGF) formalism \cite{knack} for estimation of the resistance of the MTJ stack. 
		\vspace{-3.5mm}

	\section{Device Characteristics}
	\vspace{-2mm}	
	\subsection{Writability}
		\vspace{-1mm}	

	Writing into ME devices is accomplished by application of appropriate voltages across the ME capacitor. An important parameter that dictates the write voltage and hence the write energy is the magneto-electric co-efficient ($\alpha_{ME}$). $\alpha_{ME}$ is the ratio of magnetic field generated per unit applied electric field \cite{determinis}. Experimentally, various ME materials have shown $\alpha_{ME}$ in the range 0.1/c to 1/c (c is speed of light) \cite{nikonov_bench}. In Fig. \ref{fig2} (a), we show a typical magnetization switching curve and in Fig. \ref{fig2} (b) we plot the switching probability as a function of voltage across the ME capacitor for different values of $\alpha_{ME}$. 
	It can be seen, ME materials with high $\alpha_{ME}$ are desirable for achieving low write energy.
		\vspace{-5mm}		
	\subsection{Readability}
	\vspace{-1mm}
	
\begin{figure}[t]
			\centering
			\vspace{-3mm}
			\includegraphics[width=3in,keepaspectratio]{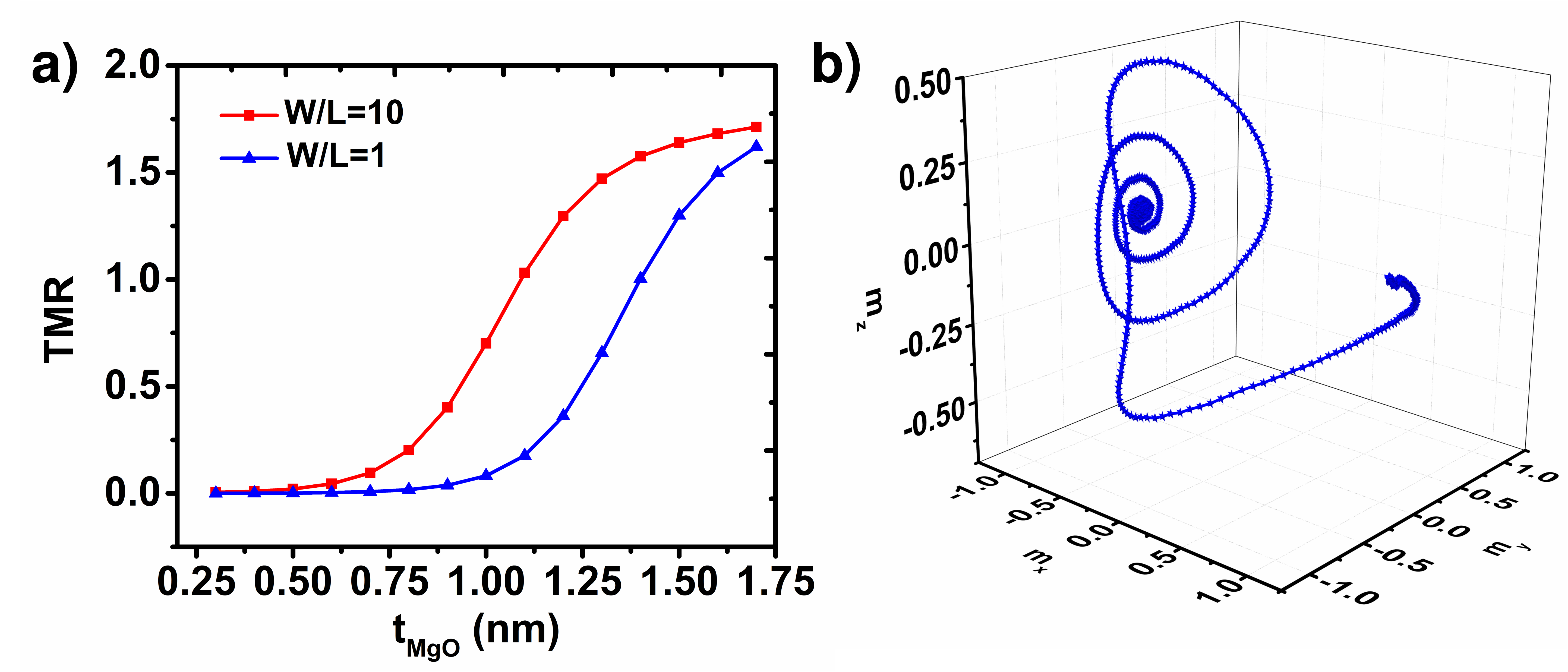}
			\vspace{-4mm}
			\caption{(a) TMR versus MgO thickness (b) A typical 3D switching trajectory }
			\vspace{-4mm}
			\label{fig3}
		\end{figure}

In a memory configuration, a CMOS transistor is used in series with the storage device. Therefore, the bit-cell TMR \textit{i.e.} the TMR of the device with the series resistance of the CMOS transistor is a more relevant metric for the sensing margin as opposed to the device TMR. In Fig. \ref{fig3}(a), we have shown the bit-cell TMR as a function of MgO thickness assuming a 45nm PTM \cite{PTM} transistor in series with varying W/L (width/length) ratios. It can be seen a higher value of MgO thickness is required to increase the bit-cell TMR and reduce the parasitic effect of the transistor series resistance \cite{amiri,jaiswal_sca}. For the ME devices, due to the decoupled read/write paths, the thickness of the MgO oxide can be increased without degrading the write efficiency (which is dictated by the ME oxide). Thus, the decoupled read/write paths for ME devices allows for better sensing due to increased bit-cell TMR.
	
		\vspace{-2mm}
	
		\vspace{-3mm}

	\subsection{Switching Speed}

Though, a detailed switching dynamics for ME devices is still under research investigation \cite{determinis}, yet it is expected that ME switching would be much faster as compared to STT switching \cite{nikonov_bench}. This is because ME switching dynamics behaves as if the magnetization direction is being switched by an external field which does not require an incubation delay  \cite{incubation} to initiate the switching process. In Fig. \ref{fig3}(b) we have shown a typical 3D trajectory of the ME switching mechanism, based on the model presented in section IV. It can be seen if the applied electric field is strong enough, the magnetization  vector starts switching without any initial incubation delay. In our simulations for an $\alpha_{ME}$ of $1/c$, complete reversal was obtained within 500ps.

		\vspace{-4mm}
	
	\section{ME Memory Design}
	
		\vspace{-1mm}
			
	\subsection{ME Dual Port Memory}

			\begin{figure}[t]
			\vspace{0mm}
			\centering
			\includegraphics[width=2.5in, keepaspectratio]{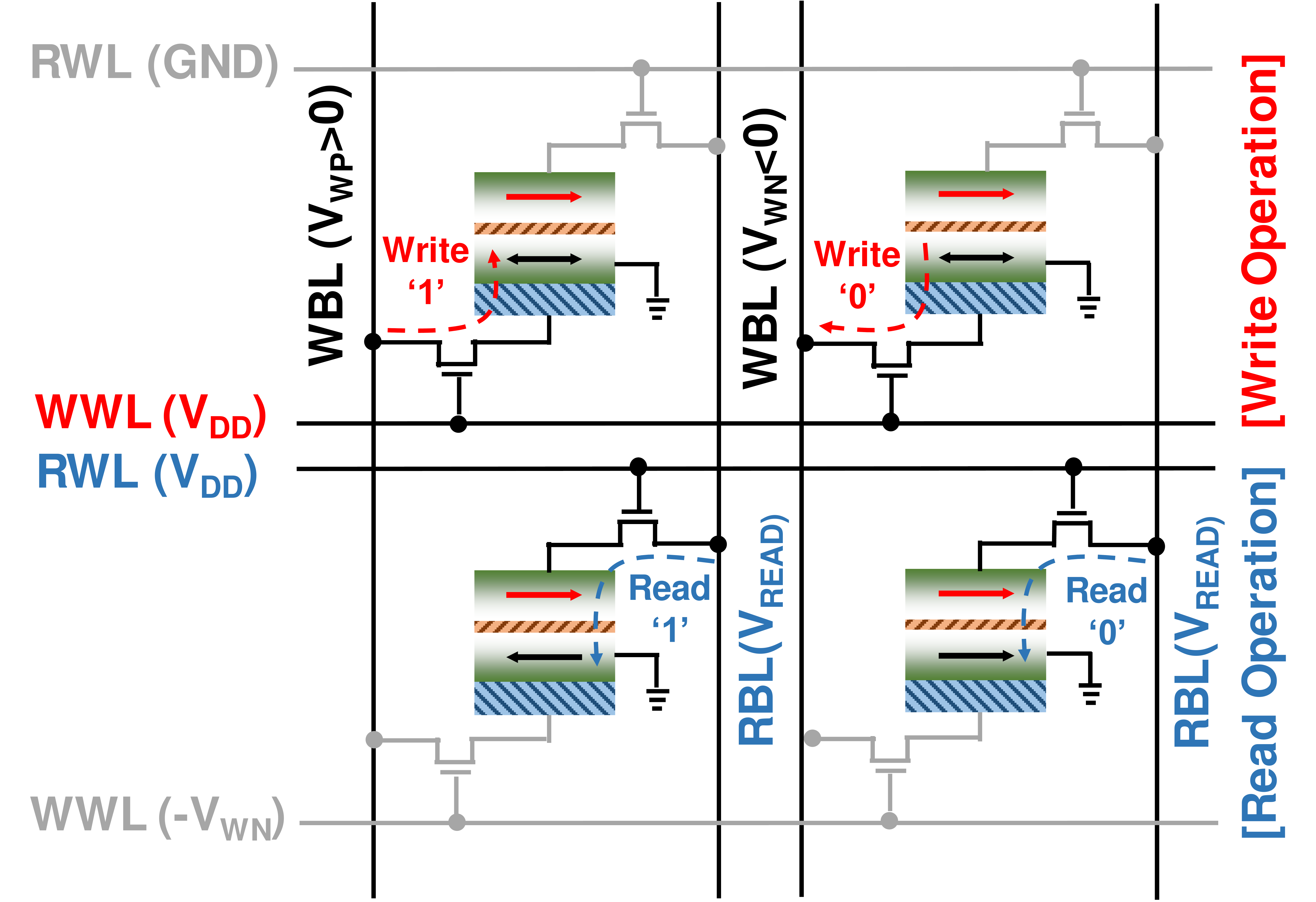}
			\vspace{-2mm}
			\caption{Dual port memory using decoupled read/write path of ME-MTJs}
			\vspace{-6mm}
			\label{fig4}
		\end{figure}
	
The proposed dual port memory using ME-MTJs is shown in Fig. \ref{fig4}. Each bit-cell consists of one ME-MTJ and two transistors. The transistor connected to WWLs are the write transistors and those connected to RWLs are the read transistors. Data can be written into the ME-MTJs by activating the write transistors of a particular row and applying appropriate write voltages (positive or negative) on WBLs. Similarly, for reading out the data, the read transistors of a given row are activated and a read voltage is applied on RBLs. The current flowing through the bit-cell is then compared with a reference to sense the current state of the ME-MTJ. 

A dual port memory is characterized by simultaneous read and write operations \textit{i.e.} while one row of the memory array is being read simultaneously another row of the memory array can be written into, thereby, improving the memory throughput \cite{yusung}. The dual port nature of the proposed ME-MTJ memory can be explained as follows. 

Let us consider row-1 in Fig. \ref{fig4} is being written into. The write transistors corresponding to row-1 would be activated and by application of proper voltages on WBLs, a P or an AP state can be written into the ME-MTJs. 
Simultaneously, the read transistors corresponding to row-2 are activated and by sensing the current flowing through the RBLs, the state of the ME-MTJs connected to row-2 can be sensed. 
Our simulations indicate, write energy consumption per bit of 0.072 fJ for $\alpha_{ME} = 1/c$ and read energy consumption of 1.3fJ for read voltage of 200mV and read time of 0.5ns. For the present proposal ME switching enables two orders of magnitude improvement in write energy and 8x improvement in switching speed as compared to STT based MTJs \cite{toshibaa}, in addition to improved TMR and throughput.

\vspace{-4mm}

	\subsection{ME CAM}

The ME-XNOR based CAM cell is shown in Fig. 5 (a). The function of M1 is to selectively provide the ME-oxide capacitor with a ground connection when Data Input Line (D$_{in}$) is activated. In the read circuit, a reference MTJ ($Ref_{MTJ}$) forms a voltage divider with the resistance of the MTJ ($R_{MTJ}$). The match signal is obtained at the drain of p-MOS M2 (denoted by node $\overline{match}$), where a low voltage indicates a match is obtained and \textit{vice-versa}. 
The node $\overline{match}$ is pre-charged to $V_{DD}$. The strengths of the n-MOS and the p-MOS transistors, connected to the $\overline{match}$ line, are adjusted such that even one activated p-MOS in a row is enough to maintain the output node in its pre-charged state.  \par
	
				\begin{figure}[t]
			\vspace{-5mm}
			\centering
			\includegraphics[width=2.5in, keepaspectratio]{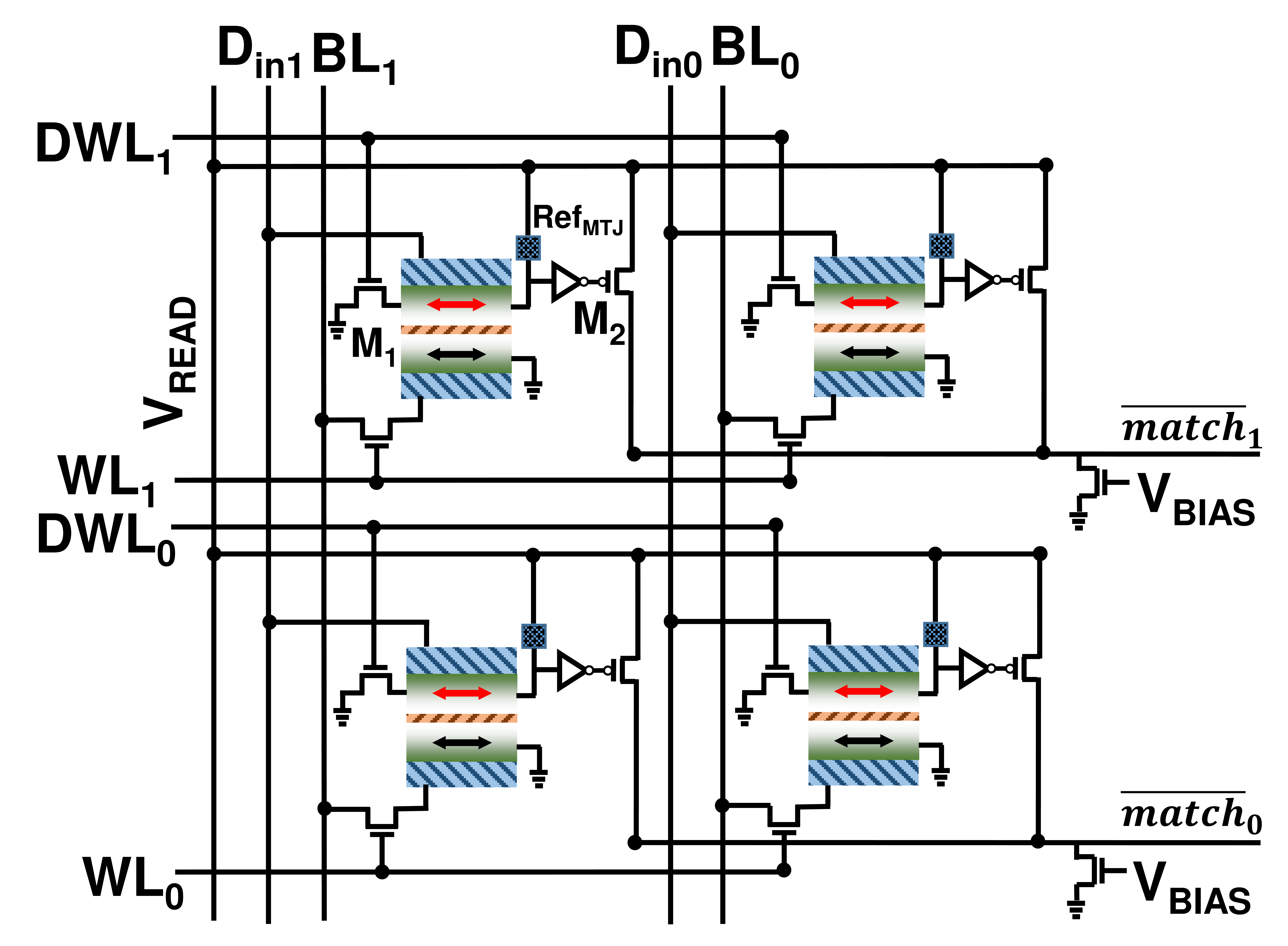}
			\vspace{-3mm}
			\caption{Proposed CAM based on ME-XNOR device.}
			\label{fig5}
			\vspace{-6mm}
		\end{figure}

	The operation of the circuit can be divided into three modes: i) Write Mode, ii) Data Input Mode and iii) Read Mode. To write data in the lower (upper) ferromagnet, a write pulse corresponding to bit `1' (positive voltage) and `0' (negative voltage), respectively, is applied on the BL (D$_{in}$) with the WL (DWL) activated. If the digital bit written in the lower ferromagnet is same as the data to be matched (stored in the upper ferromagnet), the MTJ switches to low resistance state. Finally in the \textit{read mode}, a read pulse of 1 V ($V_{READ}$) is applied for the read process. The output of the inverter goes `high' only if the MTJ is in low resistance state indicating that the bit written in the top magnet in mode (ii) matches the bit stored in the bottom magnet. Matching of all bits in a row turns all the p-MOS OFF and $\overline{match}$ goes low, indicating that a match is found. The write and read energy per bit was found to be 0.072 fJ and 15 fJ, respectively, indicating two orders of magnitude improvement in write energy and comparable read energy as compared to previous works as in \cite{vcma_cam}.

\vspace{-3mm}
	
	\section{Conclusion}
The prospects of achieving voltage driven switching of magnetization has renewed the interest for future low-power non-volatile spintronic memories. In this paper, we first analyze the writability, readability and switching speed of devices based on ME effect. Further, we propose two energy efficient memories using the ME devices. The proposed dual port memory allows for energy-efficient write operations in addition to faster speed, improved TMR and throughput. The proposed CAM requires lesser number of transistors due to the compact XNOR operation enabled by the ME XNOR device, resulting in an area-efficient as well as energy-efficient CAM.

	
	%
	
	\vspace{-4mm}
	
	

	
	\newpage


	
	
	\bibliographystyle{IEEEtran}

	\end{document}